\documentclass{aa}

\usepackage{graphicx}
\usepackage[varg]{txfonts}
\usepackage{natbib}
\usepackage[pdftex,colorlinks=true,linkcolor=blue,citecolor=blue,urlcolor=blue]{hyperref}
\usepackage{amsmath}
\usepackage{bbold}
\usepackage{tikz}
\usetikzlibrary{decorations.pathreplacing,math}

\def\expo#1{\mathbf{e}^{#1}}
\def\d{\mathrm{d}}

\def\H{\mathcal{H}}
\def\K{\mathcal{K}}

\def\t{^\mathrm{T}}
\def\mt{^\mathrm{-T}}
\def\primet{^\mathrm{\prime T}}
\def\id{\mathbb{1}}
\def\E{\mathbb{E}}
\def\R{\mathbb{R}}

\def\O{\mathcal{O}}
\def\mean#1{\overline{#1}}
\def\varmean#1{\langle#1\rangle}

\DeclareMathOperator{\tr}{tr}
\DeclareMathOperator{\diag}{diag}

\DeclareMathOperator{\sinc}{sinc}
\DeclareMathOperator{\var}{var}
\DeclareMathOperator{\cov}{cov}
\def\hadprod{\!*\!}

\begin{document}

\title{Efficient modeling of correlated noise}
\subtitle{I. Statistical significance of periodogram peaks}

\author{J.-B. Delisle\inst{1}
  \and N. Hara\inst{1,}\thanks{NCCR CHEOPS fellow}
  \and D. Ségransan\inst{1}
}
\institute{Département d'astronomie, Université de Genève,
  51 chemin des Maillettes, 1290 Versoix, Suisse\\
  \email{jean-baptiste.delisle@unige.ch}
}

\date{\today}

\abstract{
  Periodograms are common tools used to search for periodic signals in unevenly spaced time series.
  The significance of periodogram peaks is often assessed using
  false alarm probability (FAP), which in most studies
  assumes uncorrelated noise and is computed using numerical methods such as bootstrapping or Monte Carlo.
  These methods have a high computational cost, especially for low FAP levels, which are of most interest.
  We present an analytical estimate of the FAP of the periodogram in the presence of correlated noise,
  which is fundamental to analyze astronomical time series correctly.
  The analytical estimate that we derive provides a very good approximation of the FAP at a much lower cost
  than numerical methods.
  We validate our analytical approach by comparing it with Monte Carlo simulations.
  Finally, we discuss the sensitivity of the method to different assumptions in the modeling of the noise.
}

\keywords{methods: data analysis -- methods: statistical -- methods: analytical -- planets and satellites: general}

\maketitle

\section{Introduction}
\label{sec:intro}

Detecting periodic signals in unevenly spaced time series
is a common problem in astronomy, which is encountered,
for instance, when searching for binaries or exoplanet companions in radial velocity,
astrometric, or photometric time series.
The Lomb-Scargle (LS) periodogram \citep{lomb_leastsquares_1976,scargle_studies_1982}
is a classical and efficient tool to search for sinusoidal signals.
The principle of the LS periodogram is to scan a wide range of frequencies
and to compare a linear sinusoidal model at a given frequency with a constant model, called the base model.
A widespread variant of the LS periodogram, called the generalized LS periodogram (GLS),
was proposed by \citet{zechmeister_generalised_2009}
where the constant is adjusted for at each frequency
\citep[see also][]{ferraz-mello_estimation_1981}.
Once the periodogram is computed, the false alarm probability (FAP) criterion
is often used to determine whether or not it supports the detection of a periodic signal.
The FAP is often estimated using numerical methods such as bootstrapping or Monte Carlo.
These methods can be computationally intensive, especially at low FAP levels.
Indeed, estimating a FAP level of $P$ numerically requires at least the computation of $10/P$ periodograms.

Several analytical formula have been proposed for the FAP \citep{scargle_studies_1982,horne_prescription_1986},
but have been subsequently contested \citep{koen_significance_1990}.
The periodogram framework was generalized in a series of works to handle more complex models
\citep{baluev_assessing_2008,baluev_accounting_2009,baluev_detecting_2013,baluev_vonmises_2013,baluev_keplerian_2015},
where rigorous and sharp analytical approximations of the FAP were provided
based on the so-called Rice formula
and previous works by \citet{davies_hypothesis_1977,davies_hypothesis_1987,davies_hypothesis_2002}.
These works allow a fast and rigorous estimation of the periodogram FAP in the presence of white noise.
However, the white noise assumption is often incorrect for astronomical time series.
Indeed, several sources of correlated noise,
such as the astronomical source itself, Earth’s atmosphere, or instrumental systematics,
could contaminate the measurements.

For instance, stellar variability has a huge impact on the detection of low-mass exoplanets
using high-precision radial velocity time series.
Indeed, at a precision better than 1~m/s,
stellar variability affects the measurements on timescales ranging
from a few minutes (for stellar oscillation p-modes),
to hours and days (for stellar granulation and super-granulation),
and even up to the star rotation period (for the effect of spots and plages).
In this context, low-mass exoplanet detection becomes a challenge \citep[e.g.,][]{queloz_planet_2001}
and proper tools have to be developed to treat correlated noise properly and to compute reliable periodograms and FAPs.
\citet{sulis_using_2016} provide an analytical FAP estimate for periodograms normalized
by the power density spectrum of the noise, in the limit of low aliasing
and in the case of evenly sampled data.
This is however, not the case of most radial velocity datasets.
\citet{baluev_planetpack_2013} provides a "suggestive generalization"
to the correlated noise case
of the FAP formula obtained in \citet{baluev_assessing_2008},
but advocates against the use of this formula since it has not been proved rigorously.

In this article,
we extend the work of \citet{baluev_assessing_2008} to account for correlated noise.
In Sect.~\ref{sec:corrperio} we define a general periodogram for an arbitrary covariance of the noise
and provide an analytical approximation of the corresponding FAP, which we validate against Monte Carlo simulations.
In Sect.~\ref{sec:wrongnoise} we provide a method to explore the sensitivity of the periodogram to the noise model.
We discuss our results in Sect.~\ref{sec:conclusion}.

\section{Significance of periodogram peaks with correlated noise}
\label{sec:corrperio}

In this section, we present a method to assess the significance of periodogram peaks (FAP) in the
correlated noise case.
We first give a general definition of the periodogram in Sect.~\ref{sec:periodogram}.
We then provide formulas to compute the corresponding FAP in Sect.~\ref{sec:fap}.
Finally, we compare this analytical FAP with the results of Monte Carlo simulations
in Sect.~\ref{sec:montecarlo}.

\subsection{General linear periodogram}
\label{sec:periodogram}

We extend
the general definition of least squares periodograms by \citet{baluev_assessing_2008}
to the correlated noise case.
Following \citet{baluev_assessing_2008},
we compare the $\chi^2$ of the residuals of a linear base model $\H$ of $p$ parameters
with enlarged linear models $\K$ of $p+d$ parameters, parameterized by the frequency $\nu$.
The base model $\H$ is written as
\begin{equation}
  \H\ :\quad m_\H(\theta_\H) = \varphi_\H\theta_\H,
  \label{eq:mh}
\end{equation}
where $\theta_\H$ is the vector of size $p$ of the model parameters,
$\varphi_\H$ is a $n\times p$ matrix,
and $n$ the number of points in the time series.
The columns of $\varphi_H$ are thus explanatory time series
that are scaled by the linear parameters $\theta_\H$.
For instance, in the case of a radial velocity time series with two different instruments
and a linear drift, the linear base model could be chosen as
\begin{equation}
  m_\H = \gamma_1 \delta_1(t) + \gamma_2 \delta_2(t) + \alpha (t-\mathrm{epoch}),
\end{equation}
where $\gamma_1$ and $\gamma_2$ are the velocity offsets of both instruments
and $\alpha$ is the slope of the linear drift.
The function $\delta_1(t)$ (respectively, $\delta_2(t)$)
is equal to one for the points taken by instrument 1 (respectively, instrument 2) and zero otherwise.
The matrix $\varphi_\H$ would thus be a $n\times 3$ matrix whose columns
would be the three explanatory time series $\varphi_\H = (\delta_1(t), \delta_2(t), (t-\mathrm{epoch}))$
and the vector of parameters would be $\theta_\H = (\gamma_1, \gamma_2, \alpha)$.

The enlarged model $\K(\nu)$ is written as
\begin{equation}
  \K(\nu)\ :\quad m_\K(\nu, \theta_\K) = \varphi_\K(\nu)\theta_\K,
  \label{eq:mk}
\end{equation}
where $\theta_\K = (\theta_\H, \theta)$ is the vector of size $p+d$ of the parameters
and $\varphi_\K(\nu) = (\varphi_H, \varphi(\nu))$ is a $n\times (p+d)$ matrix
whose $p$ first columns are those of $\varphi_\H$,
and whose $d$ last columns are functions of the frequency $\nu$.
Typically, $d=2$, and the two additional columns are $\cos(\nu t)$ and $\sin(\nu t)$,
but the theory developed by \citet{davies_hypothesis_1977,davies_hypothesis_1987,davies_hypothesis_2002}
and \citet{baluev_assessing_2008} is more general.

We denote by $\chi_\H^2$ and $\chi_\K^2(\nu)$ the $\chi^2$ of the residuals
after a linear least squares fit
with a covariance matrix $C$
of the models $\H$ and $\K(\nu)$, respectively.
\citet{baluev_assessing_2008} assumed the noise to be independent (diagonal covariance matrix)
and Gaussian and that the uncertainties of the measurements
are known precisely (at least within a common factor).
In this generalization,
we do not assume the noise to be independent anymore,
but we still assume the noise to be Gaussian with a known
covariance matrix $C$ (at least within a common factor).
The covariance matrix $C$ accounts for all sources of correlated and uncorrelated noise, such as
intrinsic noise from the source or subsequent contamination by the Earth’s atmosphere or by the instrument.

In the general case, the periodogram is a function $z(\nu) = f(\chi_\H^2, \chi_\K^2(\nu))$.
A general linear periodogram $z$ is thus defined by the models $\H$ and $\K(\nu)$ and the function $f$.
In the following, we consider the four definitions of the periodogram proposed by \citet{baluev_assessing_2008}:
\begin{align}
  \label{eq:defperio}
  z_0(\nu) = \frac{1}{2}\left(\chi^2_\H-\chi^2_\K(\nu)\right), \qquad &
  z_1(\nu) = \frac{n_\H}{2} \frac{\chi_\H^2 - \chi_\K^2(\nu)}{\chi_\H^2},\nonumber\\
  z_2(\nu) = \frac{n_\K}{2} \frac{\chi_\H^2 - \chi_\K^2(\nu)}{\chi_\K^2(\nu)}, \qquad &
  z_3(\nu) = \frac{n_\K }{2} \ln \frac{\chi_\H^2}{\chi_\K^2(\nu)},
\end{align}
where $n_\H = n - p$ and $n_\K = n - (p + d)$.

The widespread GLS periodogram
\citep[see][]{ferraz-mello_estimation_1981,zechmeister_generalised_2009}
is very close to the definition $z_1$ of the periodogram.
Indeed, we have
\begin{equation}
  \label{eq:GLS}
  z_\mathrm{GLS} = \frac{\chi_\H^2 - \chi_\K^2(\nu)}{\chi_\H^2} = \frac{2}{n_\H} z_1(\nu),
\end{equation}
and all the results obtained for $z_1$ are also valid for the GLS.

Once the periodogram is computed, it is useful to compute the $p$-value of the highest peak,
or FAP, defined as $\mathrm{Pr}\{\max_\nu z(\nu) \geqslant Z\ |\ \H\}$,
where $Z$ is the value of the maximum peak of the periodogram computed on the data.

\subsection{False alarm probability for periodograms with correlated noise}
\label{sec:fap}

\begin{table}
  \caption{False alarm probability for different definitions (see Eq.~(\ref{eq:defperio}))
    of the periodogram power by \citet{baluev_assessing_2008}
    in the case $d=2$.}
  \begin{center}
    \begin{tabular}{l|ll}
      \hline
      \hline
      $z(\nu)$ & $\mathrm{FAP_{single}}(Z)$ & $\tau(Z, \nu_\mathrm{max})$, approximately\\
      \hline
      $z_0(\nu)$ &
      $\expo{-Z}$ &
      $W\expo{-Z}\sqrt{Z}$ \\
      $z_1(\nu)$ &
      $\left(1-\frac{2Z}{n_\H}\right)^\frac{n_\K}{2}$ &
      $\gamma_\H W\left(1-\frac{2Z}{n_\H}\right)^\frac{n_\K-1}{2}\sqrt{Z}$ \\
      $z_2(\nu)$ &
      $\left(1+\frac{2Z}{n_\K}\right)^{-\frac{n_\K}{2}}$ &
      $\gamma_\K W\left(1+\frac{2Z}{n_\K}\right)^{-\frac{n_\K}{2}}\sqrt{Z}$ \\
      $z_3(\nu)$ &
      $\expo{-Z}$ &
      $\gamma_\K W\expo{-Z\left(1-\frac{1}{2 n_\K}\right)}\sqrt{n_\K\sinh\frac{Z}{n_\K}}$ \\
      \hline
    \end{tabular}
  \end{center}
  \tablefoot{Factor $W$ is the rescaled frequency bandwidth
    defined in Eq.~(\ref{eq:defW}) and $\Gamma$ is Euler's gamma function.
    The factors $\gamma_{\H,\K} = \sqrt{\frac{2}{n_{\H,\K}}} \Gamma(\frac{n_\H}{2})/\Gamma(\frac{n_\H-1}{2})$
    can be neglected for $n_\H \geq 10$. }
  \label{tab:power}
\end{table}

In this section, we provide analytical approximations of the FAP
for the definitions of the periodogram of Eq.~(\ref{eq:defperio}).
Their precise derivation is provided in Appendix~\ref{sec:fapcomputation}.

The model $\H$ is defined as in Eq.~(\ref{eq:mh}),
where the $n \times p$ matrix $\varphi_\H$ is user defined;
for instance, it might include offsets and drifts.
The model $\K$ (eq.~(\ref{eq:mk})) is the horizontal concatenation of $\varphi_\H$
and the two column vectors $\cos\nu t$ and $\sin\nu t$
($\varphi_\K(\nu) = (\varphi_\H, \cos\nu t,  \sin\nu t)$).

The periodogram is computed in the range of frequencies $]0, \nu_\mathrm{max}]$.
The FAP is approximated by
\citep[see][]{baluev_assessing_2008}
\begin{equation}
  \label{eq:FAPmax}
  \mathrm{FAP_{max}}(Z, \nu_\mathrm{max}) \approx 1 - \left(1-\mathrm{FAP_{single}}(Z)\right)\expo{-\tau(Z, \nu_\mathrm{max})},
\end{equation}
where analytical expressions for $\mathrm{FAP_{single}}$ and $\tau(Z, \nu_\mathrm{max})$
are given in Table~\ref{tab:power}.
These expressions depend on the rescaled frequency bandwidth $W$ defined as
\begin{equation}
  \label{eq:defW}
  W = \frac{\nu_\mathrm{max}}{2\pi} T_\mathrm{eff},
\end{equation}
where $T_\mathrm{eff}$ is the effective time series length,
which we approximate by (see Appendix~\ref{sec:analyticalfap})
\begin{equation}
  \label{eq:Teffestimate}
  T_\mathrm{eff} \approx \sqrt{4\pi}
  \sqrt{\frac{\varmean{\Pi\hadprod\sinc \nu_\mathrm{max}\Delta}}{\varmean{\sinc \nu_\mathrm{max}\Delta}}
    - \left(\frac{\varmean{\Sigma\hadprod\sinc \nu_\mathrm{max}\Delta}}{2{\varmean{\sinc \nu_\mathrm{max}\Delta}}}\right)^2}.
\end{equation}
The $n \times n$ matrices $\Sigma$, $\Delta$, and $\Pi$ are defined as
\begin{align}
  \Sigma_{i,j} & = t_i + t_j,\nonumber \\
  \Delta_{i,j} & = t_i - t_j,\nonumber \\
  \Pi_{i,j}    & = t_i t_j,
\end{align}
and for two $n\times n$ matrices $X$ and $Y$,
$X \hadprod Y$ is the Hadamard (or element-wise) product
\begin{equation}
  (X \hadprod Y)_{i,j} = X_{i,j} Y_{i,j},
\end{equation}
and $\varmean{X}$ is defined as
\begin{equation}
  \varmean{X} = \sum_{i,j} C^{-1}_{i,j} X_{i,j}.
\end{equation}
The expression of the effective time series length found by \citet{baluev_assessing_2008} in the white noise case
can be derived from Eq.~(\ref{eq:Teffestimate}).
Indeed in this case (diagonal covariance matrix $C$), Eq.~(\ref{eq:Teffestimate}) is simplified as
\begin{equation}
  \label{eq:Teffwhite}
  T_\mathrm{eff} \approx \sqrt{4\pi\left(\mean{t^2}-\mean{t}^2\right)},
\end{equation}
where $\mean{t}$, and $\mean{t^2}$ are weighted means with weights $C^{-1}_{i,i}/\sum_j C^{-1}_{j,j}$.

In Appendix~\ref{sec:analyticalfap},
we additionally provide approximations of $T_\mathrm{eff}$ in the low and high frequency limit,
that is, $\nu_\mathrm{max}\Delta \ll 1$ (Eq.~(\ref{eq:Tefflow}))
and $\nu_\mathrm{max}\Delta \gg 1$ (Eq.~(\ref{eq:Teffhigh})).

\subsection{Comparison of analytical FAP with Monte Carlo simulations}
\label{sec:montecarlo}

\begin{table}
  \caption{Values of the covariance matrix parameters (see Eq.~(\ref{eq:Cij}))
  for the four noise models used in our study of the \object{HD~136352} radial velocity time series.}
  \begin{center}
    \begin{tabular}{c|cccc}
      \hline
      \hline
      & obs. & jit. & daily exp. & monthly exp. \\
      \hline
      $\sigma_\mathrm{jit.}$ (m/s) & -- & 1 & -- & -- \\
      $\sigma_\mathrm{exp.}$ (m/s) & -- & -- & 1 & 1 \\
      $\tau_\mathrm{exp.}$ (d) & -- & -- & 1 & 30 \\
      \hline
    \end{tabular}
  \end{center}
  \label{tab:noiseparams}
\end{table}

\begin{figure}
  \centering
  \includegraphics[width=0.75\linewidth]{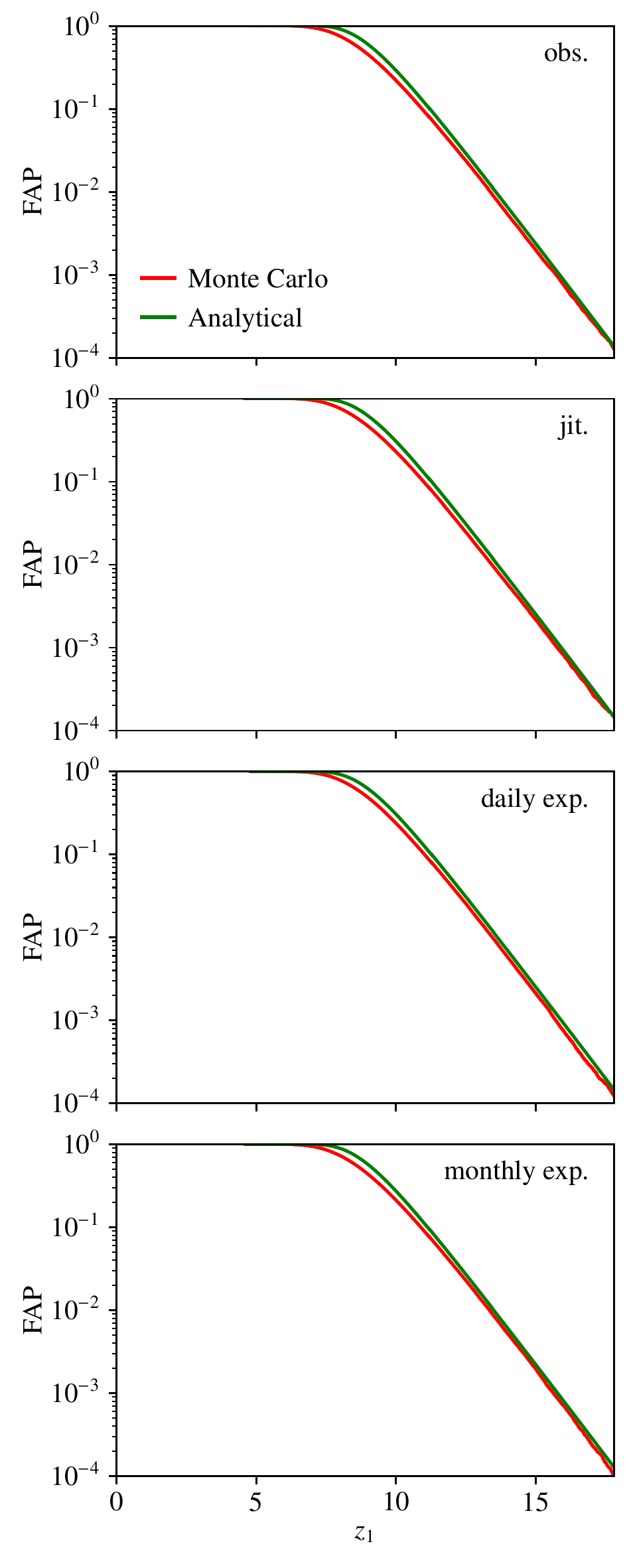}
  \caption{Comparison between analytical and numerical estimations of the FAP for four types of covariance matrix
  (see Sect.~\ref{sec:montecarlo}) and using the HARPS time series of \object{HD~136352}.
  The periodogram power is computed following the definition $z_1$ of Eq.~(\ref{eq:defperio}).
  The expectation of $z_1$ is one (see Eq.~(\ref{eq:expectperiotrue})).}
  \label{fig:compFAP_HD136352}
\end{figure}

We validated our analytical estimate of the FAP
(Eqs.~(\ref{eq:FAPmax})-(\ref{eq:Teffestimate}), Table~\ref{tab:power})
by comparing it with Monte Carlo simulations.
The Monte Carlo simulations are performed by generating a large set of random time series following the same distribution
(same covariance matrix).
We used the times of observation and error bars of the HARPS radial velocities
of \object{HD~136352} \citep{udry_harps_2019}
to obtain a realistic temporal sampling and realistic covariance matrices.
The HARPS radial velocities of \object{HD~136352} consist of 648 points
taken over almost 11 years (2004-2015)
and spread over 238 distinct nights (about 2.7 points per night).

Our method is general and does not require a particular shape for the covariance matrix.
However, for illustration purposes, we assume the covariance matrix to follow
\begin{equation}
  \label{eq:Cij}
  C_{i,j} = \delta_{i,j} (\sigma_i^2 + \sigma_\mathrm{jit.}^2) + \sigma_\mathrm{exp.}^2 \expo{-|t_i-t_j|/\tau_\mathrm{exp.}}
\end{equation}
and vary the values of the parameters ($\sigma_\mathrm{jit.}$, $\sigma_\mathrm{exp.}$, $\tau_\mathrm{exp.}$)
to define four different noise models:
\begin{enumerate}
  \item obs. (white noise): a diagonal matrix with observational error bars;
  \item jit. (white noise): a diagonal matrix with observational error bars plus a jitter of 1~m/s;
  \item daily exp. (correlated noise): observational error bars on the diagonal, plus an exponential decay of 1~m/s with a timescale of 1~d;
  \item monthly exp. (correlated noise): the same as daily exp. but with a timescale of 30~d.
\end{enumerate}
The values of the parameters ($\sigma_\mathrm{jit.}$, $\sigma_\mathrm{exp.}$, $\tau_\mathrm{exp.}$)
used for each noise model are summarized in Table~\ref{tab:noiseparams}.
In the context of radial velocity time series, jitter terms might model both intrinsic noise from the star and instrumental noise,
while exponential decay terms are often used to account for stellar noise (e.g., granulation and oscillation).

For a given covariance matrix $C$,
we generated a synthetic radial velocity time series by randomly sampling from a normal distribution
with covariance matrix $C$.
We generate $10^6$ such random time series and compute a periodogram (with the correct covariance matrix)
for each time series.
The periodograms are computed in the range $]0,\frac{2\pi}{P_\mathrm{min}}]$
where $P_\mathrm{min}=0.9~\mathrm{d}$,
and with an instrumental offset $\gamma$ adjusted for each frequency ($p=1$, $n_\H=n-1$, $n_\K=n-3$).
Then, the distribution of the maximum of these periodograms allows us to estimate numerically the FAP.

The comparison between the analytical and numerical FAP is shown
in Fig.~\ref{fig:compFAP_HD136352}.
As explained by \citet{baluev_assessing_2008}, the analytical formula of the FAP
is an upper bound that asymptotically (for low FAP levels) converges to the exact FAP.
This is indeed what we observe in Fig.~\ref{fig:compFAP_HD136352}.
For all the covariance matrices,
the analytical and numerical estimates agree very well for $\mathrm{FAP}\lesssim 0.1$,
and the analytical formula overestimates the FAP for $\mathrm{FAP} \gtrsim 0.1$

Since we used $10^6$ samples for the numerical estimation of the FAP,
we could not reliably explore FAP levels below $10^{-4}$ owing to small numbers statistics.
However, the analytical and numerical estimates agree very well down to $10^{-4}$,
and the analytical approximation is expected to be even more accurate for lower FAP levels.

To sum up, the analytical estimate provides a very good approximation of the FAP
in the range of most interest ($\mathrm{FAP}\lesssim 0.1$),
and is conservative for higher FAP levels.
Moreover, this analytical estimate is much faster to compute
than Monte Carlo simulations (or other numerical methods),
especially for low FAP levels.
These properties make it very convenient to use in practical applications.

\section{Sensitivity of periodogram to noise model}
\label{sec:wrongnoise}

\begin{figure*}
  \centering
  \includegraphics[width=\linewidth]{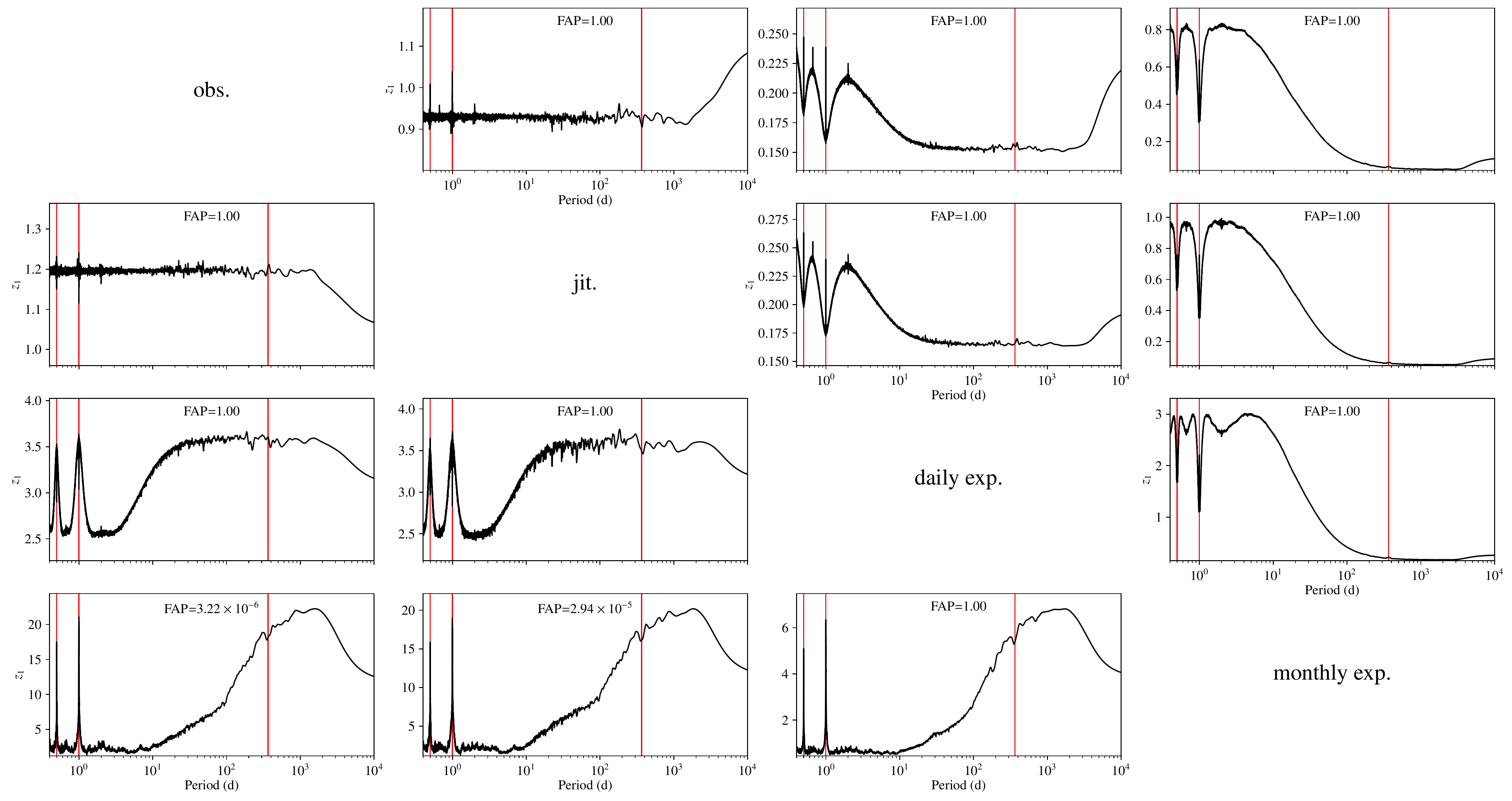}
  \caption{Periodogram expectation for \object{HD~136352} in the case of a wrong noise model
  (see Eq.~(\ref{eq:expectperioO1})).
  Rows correspond to the true noise model, while columns correspond to the assumed wrong model.
  The definition $z_1$ of the periodogram power is used (see Eq.~\ref{eq:defperio}).
  If the noise model was correct, the expectation of the periodogram power would be uniformly 1
  (see Eq.~(\ref{eq:expectperiotrue})).
  The red vertical lines highlight 0.5~d, 1~d, and 1~yr.
  For each periodogram, we provide the analytical estimate of the FAP (using the wrong noise model)
  corresponding to the highest peak of the periodogram expectation.}
  \label{fig:perio_HD136352}
\end{figure*}

The FAP formula obtained in Sect.~\ref{sec:fap} provides an efficient and
robust way to assess the significance of periodogram peaks
when the covariance matrix is known.
However, this is not the case in general,
and we can often only make educated guesses about the shape of the covariance matrix.
It is therefore necessary to explore the sensitivity of the periodogram to the noise model.
To do so, we computed the expectation of the periodogram computed with an incorrect noise model.
The analytical derivation of the periodogram expectation is described in Appendix~\ref{sec:wrongnoiseformula}.
In this section, we illustrate its use by exploring the effect of an incorrect noise model
on the periodogram and its associated FAP.

We assume that the actual covariance matrix of the noise is $C$,
while the periodogram is computed with an incorrect covariance matrix $V$.
We used the same dataset (HARPS radial velocities of \object{HD~136352})
and the same noise models as in Sect.~\ref{sec:montecarlo}.
We first chose a noise model
among the four models of Sect.~\ref{sec:montecarlo} (obs., jit., daily exp., and monthly exp.),
which we considered the correct noise model (covariance matrix $C$).
We then chose the incorrect noise model (covariance matrix $V$) among the three other models.
The study is done with the definition $z_1$ of the periodogram (see Eq.~(\ref{eq:defperio})).
We first computed the periodogram expectation following the analytical expression of Eq.~(\ref{eq:expectperioO1})
and we then estimated the FAP corresponding to the highest peak of this mean periodogram
using the analytical formula of Eq.~(\ref{eq:FAPmax}) with the incorrect covariance matrix ($V$).

The results are shown in Fig.~\ref{fig:perio_HD136352} for all pairs of noise models.
We observe in Fig.~\ref{fig:perio_HD136352}
that adding a jitter term (jit.) does not affect the results much compared
to the model with the observational error bars alone (obs.),
and vice versa.
This is not surprising since all the data points were taken with the same instruments (HARPS),
and thus have very similar error bars.
Therefore, adding a common jitter term to all error bars does not much affect the relative weight
of each measurement.
Moreover, the definition $z_1$ of the periodogram (Eq.~(\ref{eq:defperio}))
is not sensitive to the multiplication of all error bars by a common factor.

Using a correlated noise model (daily or monthly exp.) while the true
noise is uncorrelated (obs. or jit.) remains conservative
on the whole range of frequencies
($\E(z_1(\nu)) \simeq 1$ for all $\nu$).
However, the periodogram level is very low at long periods,
which means that the detection capability at long periods is strongly reduced.
On the other hand, using an uncorrelated noise model (obs. or jit.)
while the true noise model is correlated (daily or monthly exp.)
can lead to spurious detections with very low FAP levels
(down to $3.22\times 10^{-6}$, see Fig.~\ref{fig:perio_HD136352}).
Underestimating the correlation timescale (using daily instead of monthly exp.)
has a similar (but weaker) effect as using an uncorrelated model instead of a correlated model.
Finally overestimating the correlation timescale (using monthly instead of daily exp.)
reduces the capability to detect long periods, and might lead to spurious detections of short periods.
In the two latter cases (underestimation and overestimation of the correlation timescale),
the FAP remains very high (close to 1, i.e., non-significant detection).

Overall these results are not surprising but illustrate the possibility to investigate
the sensitivity of the periodogram and its FAP with respect to the noise model using
the formula for the periodogram expectation (Eq.~(\ref{eq:expectperioO1})).

\section{Conclusions}
\label{sec:conclusion}

We present a generalization of the analytical estimate of \citet{baluev_assessing_2008}
(which was restricted to the white noise case)
to the correlated noise case (see Sect.~\ref{sec:fap}).
We show that the "suggestive generalization" of \citet{baluev_planetpack_2013}
is valid in the low frequency limit (see Eq.~(\ref{eq:Tefflow})),
but we find a more general expression (Eq.~(\ref{eq:Teffestimate})) that is valid for all frequencies.
We validate our analytical estimate against Monte Carlo simulations (see Sect.~\ref{sec:montecarlo})
and show that this analytical criterion is very efficient and accurate,
provided that the covariance matrix of the noise is known (at least within a common factor).

In most cases, however, astronomical time series are contaminated by
sources of correlated noise that are difficult to characterize fully,
which results in an approximate modeling of the covariance matrix.
We illustrate the sensitivity of the periodogram to the noise model,
by deriving the expectation of a periodogram computed with an incorrect covariance matrix
(see Sect.~\ref{sec:wrongnoise}).
This method allows us to visualize which parts of the periodogram are the most affected
by a change in the noise model.
For instance, we observe, as expected,
that overestimating the correlation timescale of the noise tends to
reduce the detection capability at long periods strongly,
while underestimating this timescale can lead to spurious detections.
Another way to visualize the sensitivity of the periodogram to the noise model
would be to compute several periodograms on the same data with various noise models.
Both approaches are complementary to better understand the features observed in the periodograms.

Several methods can be used to obtain a more realistic covariance matrix.
First, a likelihood maximization can be performed to adjust some noise model parameters.
This maximization can be performed once,
before computing the least squares periodogram (with a fixed noise model).
It can also be performed for each frequency by computing a likelihood periodogram
instead of a least squares periodogram.
In the white noise case, \citet{baluev_accounting_2009}
proposed a FAP estimate for a likelihood periodogram with a free error term (jitter)
added in quadrature to the nominal error bars and adjusted for at each frequency.
A generalization to the correlated noise case could probably also be achieved
for this likelihood periodogram, but is beyond the scope of this article.
Finally, a Bayesian approach could be used to compare different models (signal + noise),
but with a much higher computational cost.

\begin{acknowledgements}
  We thank the anonymous referee for his/her useful comments.
  We acknowledge financial support from the Swiss National Science Foundation (SNSF).
  This work has, in part, been carried out within the framework of
  the National Centre for Competence in Research PlanetS
  supported by SNSF.
\end{acknowledgements}

\bibliographystyle{aa}
\bibliography{fap}

\appendix

\section{Computation of the FAP in the correlated noise case}
\label{sec:fapcomputation}

In this appendix, we extend the method of \citet{baluev_assessing_2008}
to obtain analytical FAP estimates in the correlated noise case.
The main idea allowing the analytical approximation of the
FAP with correlated noise is to perform a change of coordinates
that yields independent Gaussian random variables.
Then, the method described by \citet{baluev_assessing_2008} can be applied on these new variables.
The change of variables is described in Sect.~\ref{sec:changevar}
and the derivation of the FAP estimate in Sect.~\ref{sec:analyticalfap}.

\subsection{Change of random variables}
\label{sec:changevar}

Let us assume that the covariance matrix $C$ of the noise is known (at least within a common factor).
Then, under the null hypothesis (i.e., assuming the base model is correct), the time series is written as
\begin{equation}
  \label{eq:yH}
  y = \varphi_\H \theta_{\H,0} + \epsilon,
\end{equation}
where $\theta_{\H,0}$ is the true value of the parameters
and the noise $\epsilon$ is Gaussian with zero mean and covariance $C$.

For a linear model $\varphi_m$ and parameters $\theta_m$ ($m=\H$ or $\K$),
the $\chi^2$ is defined as
\begin{align}
  \chi^2(\theta_m) & = (y-\varphi_m\theta_m)\t C^{-1} (y-\varphi_m\theta_m)\nonumber      \\
                   & = (\varphi_\H \theta_{\H,0} - \varphi_m\theta_m + \epsilon)\t C^{-1}
  (\varphi_\H \theta_{\H,0} - \varphi_m\theta_m + \epsilon).
\end{align}
The least squares estimate of the parameters is written as
\begin{align}
  \hat{\theta}_m & = \left(\varphi_m\t C^{-1} \varphi_m\right)^{-1} \varphi_m\t C^{-1} y\nonumber                \\
                 & = \theta_{\H,0} + \left(\varphi_m\t C^{-1} \varphi_m\right)^{-1} \varphi_m\t C^{-1} \epsilon,
\end{align}
and the minimum $\chi^2$ is thus
\begin{align}
  \chi^2_m & = \min_{\theta_m} \chi^2(\theta_m) = \chi^2(\hat{\theta}_m) = (y-\varphi_m\hat{\theta})\t C^{-1} (y-\varphi_m\hat{\theta})\nonumber \\
           & = y\t \left(C^{-1} - C^{-1} \varphi_m \left(\varphi_m\t C^{-1} \varphi_m\right)^{-1} \varphi_m\t C^{-1}\right) y\nonumber           \\
           & = \epsilon\t \left(C^{-1} - C^{-1} \varphi_m \left(\varphi_m\t C^{-1} \varphi_m\right)^{-1} \varphi_m\t C^{-1}\right) \epsilon.
\end{align}
Let us now perform the following change of coordinates:
\begin{align}
  \label{eq:changevar}
  z      & = L^{-1} y,\nonumber        \\
  \eta   & = L^{-1} \epsilon,\nonumber \\
  \psi_m & = L^{-1} \varphi_m,
\end{align}
where $C=L L\t$ is the Cholesky decomposition of the covariance matrix.
Since we assumed $\epsilon$ to follow a Gaussian law with zero mean and covariance $C$,
$\eta$ follows a Gaussian law with zero mean and covariance $\id$.
The random variables $\eta$ are thus independent Gaussian variables.
In these new variables, the $\chi^2$ is simply rewritten as
\begin{equation}
  \chi^2(\theta) = (z-\psi_m\theta_m)\t (z-\psi_m\theta_m),
\end{equation}
the least squares estimate of the parameters is rewritten as
\begin{equation}
  \hat{\theta}_m = \theta_{\H,0} + \left(\psi_m\t \psi_m\right)^{-1} \psi_m\t \eta,
\end{equation}
and the minimum $\chi^2$ as
\begin{equation}
  \chi^2_m = \eta\t \left(\id - \psi_m \left(\psi_m\t \psi_m\right)^{-1} \psi_m\t\right) \eta,
\end{equation}
which follows a $\chi^2$ law with $n_m$ degrees of freedom ($n_\H = n-p$, $n_\K = n-(p+d)$).
Therefore, the initial problem of analyzing a time series $y$
with covariance matrix $C$,
base model $\varphi_\H$, and enlarged models $\varphi_\K = (\varphi_H, \varphi(\nu))$
is equivalent to analyzing the time series $z$, with covariance matrix $\id$,
base model $\psi_\H$, and enlarged model $\psi_\K = (\psi_H, \psi(\nu))$.
However, if $\varphi(\nu)$ was the sine and cosine at frequency $\nu$,
this is no longer the case for $\psi(\nu)$.
Nevertheless, the theory of \citet{baluev_assessing_2008} is very general,
and does not assume a particular shape for this matrix, except for the final application
to the least squares periodogram.
We thus follow the method proposed by \citet{baluev_assessing_2008},
and only change the hypothesis on the shape of the enlarged model matrix.

\subsection{Analytical FAP estimate}
\label{sec:analyticalfap}

The FAP can be bounded by \citep[see][Eq.~(5)]{baluev_assessing_2008}
\begin{equation}
  \mathrm{FAP_{max}}(Z, \nu_\mathrm{max}) \leq \mathrm{FAP_{single}}(Z) + \tau(Z, \nu_\mathrm{max}),
\end{equation}
and approximated by \citep[see][Eq.~(6)]{baluev_assessing_2008}
\begin{equation}
  \mathrm{FAP_{max}}(Z, \nu_\mathrm{max}) \approx 1 - \left(1-\mathrm{FAP_{single}}(Z)\right)\expo{-\tau(Z, \nu_\mathrm{max})},
\end{equation}
where $Z$ is the maximum periodogram power,
$\mathrm{FAP_{single}}(Z)$ is the FAP in the case in which the frequency $\nu$
of the putative additional signal is fixed,
$\tau(Z, \nu_\mathrm{max})$ is the expectation of the number of up-crossings of the level $Z$
by the periodogram \citep[see][]{baluev_assessing_2008}.

Computing $\mathrm{FAP_{single}}(Z)$ and $\tau(Z, \nu_\mathrm{max})$
requires us to specify the definition of the periodogram $z(\nu)$.
\citet{baluev_assessing_2008} proposed several definitions
and derived the corresponding formulas for $\mathrm{FAP_{single}}(Z)$ and $\tau(Z, \nu_\mathrm{max})$.
These results are summarized in Table~\ref{tab:power} for $d=2$.
The general case ($d$ not necessarily equal to 2 and other definitions of $z(\nu)$)
is provided in \citet{baluev_assessing_2008}, Appendix~B.

For the definitions of the periodogram of Eq.~(\ref{eq:defperio}) and assuming $d=2$,
the only quantity left to compute is the factor $W$,
which is the rescaled frequency bandwidth, defined as \citep[see][]{baluev_assessing_2008}
\begin{equation}
  W = \frac{A(\nu_\mathrm{max})}{2\pi^{3/2}},
\end{equation}
where
\begin{align}
  \label{eq:Anumax}
  A(\nu_\mathrm{max}) & = \int_{0}^{\nu_\mathrm{max}}\d\nu \int_{x^2<1} \frac{\sqrt{x\t M(\nu) x}}{x\t x} \d x\nonumber \\
                      & \leq 2\pi \int_{0}^{\nu_\mathrm{max}} \sqrt{\frac{\tr(M(\nu))}{2}} \d \nu.
\end{align}
The $2\times 2$ matrix $M(\nu)$ is defined as follows (with $x' = \partial x/\partial\nu$):
\begin{align}
  \label{eq:defM}
  & Q = \psi\t \psi = \varphi\t C^{-1} \varphi, \qquad
  S = \psi\t \psi' = \varphi\t C^{-1} \varphi',\nonumber                            \\
  & R = \psi\primet \psi' = \varphi\primet C^{-1} \varphi',\nonumber               \\
  & Q_\H = \psi_\H\t \psi = \varphi_\H\t C^{-1} \varphi, \qquad
  S_\H = \psi_\H\t \psi' = \varphi_\H\t C^{-1} \varphi',\nonumber                   \\
  & Q_{\H,\H} = \psi_\H\t \psi_\H = \varphi_\H\t C^{-1} \varphi_\H,\nonumber       \\
  & \tilde{Q} = Q - Q_\H\t Q_{\H,\H}^{-1} Q_\H, \qquad
  \tilde{S} = S - Q_\H\t Q_{\H,\H}^{-1} S_\H,\nonumber                              \\
  & \tilde{R} = R - S_\H\t Q_{\H,\H}^{-1} S_\H,\nonumber                           \\
  & M = \tilde{Q}^{-1}\left(\tilde{R} - \tilde{S}\t\tilde{Q}^{-1}\tilde{S}\right).
\end{align}
\citet{baluev_assessing_2008} also defined the effective time series length as
\begin{equation}
  \label{eq:defTeff}
  T_\mathrm{eff} = \frac{A(\nu_\mathrm{max})}{\sqrt{\pi}\nu_\mathrm{max}},
\end{equation}
such that
\begin{equation}
  W = \frac{\nu_\mathrm{max}}{2\pi} T_\mathrm{eff}.
\end{equation}
From Eqs.~(\ref{eq:Anumax}) and (\ref{eq:defTeff}), we obtain
\begin{equation}
  \label{eq:Teff}
  T_\mathrm{eff} = \frac{1}{\sqrt{\pi}}\mean{\int_{x^2<1} \frac{\sqrt{x\t M(\nu) x}}{x\t x} \d x}
  \leq \mean{\sqrt{2\pi\tr(M(\nu))}},
\end{equation}
where $\mean{x}$ is the mean of $x(\nu)$ over the frequency range $]0,\nu_\mathrm{max}]$.
As noted by \citet{baluev_assessing_2008}, the inequality in Eqs.~(\ref{eq:Anumax}) and (\ref{eq:Teff})
is very sharp in practical applications.
In particular, it saturates (i.e., becomes an equality) when the eigenvalues of $M(\nu)$ are equal.
This expression can be evaluated numerically by sampling the frequency over the interval $]0,\nu_\mathrm{max}]$,
and computing $M(\nu)$ according to Eq.~(\ref{eq:defM}) for each frequency $\nu$.
The cost of evaluating $T_\mathrm{eff}$ is of the same order of magnitude as computing the
periodogram itself.
It is therefore much more efficient than performing Monte Carlo simulations.
However, this cost is not negligible compared to the periodogram cost
and analytical approximations might be useful.

We now specify the expression of $T_\mathrm{eff}$ for  $\varphi = (\cos(\nu t), \sin(\nu t))$.
Replacing $\varphi$
in the definitions of $Q$, $S$, and $R$,
we find
\begin{align}
  Q & = \frac{1}{2}\left(\arraycolsep=2.0pt\begin{array}{cc}
      \varmean{\cos\nu\Sigma + \cos\nu\Delta} &
      \varmean{\sin\nu\Sigma}                   \\
      \varmean{\sin\nu\Sigma}                 &
      \varmean{\cos\nu\Delta - \cos\nu\Sigma}
    \end{array}\right),\nonumber \\
  S & = \frac{1}{4}\left(\arraycolsep=2.0pt\begin{array}{cc}
      -\varmean{\Sigma\hadprod\sin\nu\Sigma + \Delta\hadprod\sin\nu\Delta} &
      \varmean{\Sigma\hadprod(\cos\nu\Sigma + \cos\nu\Delta)}                \\
      \varmean{\Sigma\hadprod(\cos\nu\Sigma - \cos\nu\Delta)}              &
      \varmean{\Sigma\hadprod\sin\nu\Sigma - \Delta\hadprod\sin\nu\Delta}
    \end{array}\right),\nonumber \\
  R & = \frac{1}{2}\left(\arraycolsep=2.0pt\begin{array}{cc}
      \varmean{\Pi\hadprod(\cos\nu\Delta - \cos\nu\Sigma)} &
      -\varmean{\Pi\hadprod\sin\nu\Sigma}                    \\
      -\varmean{\Pi\hadprod\sin\nu\Sigma}                  &
      \varmean{\Pi\hadprod(\cos\nu\Sigma + \cos\nu\Delta)}
    \end{array}\right),
\end{align}
where
$\,\hadprod\,$ denotes the Hadamard (or elementwise) product,
\begin{align}
  \varmean{X}  & = \sum_{i,j} C^{-1}_{i,j} X_{i,j},\nonumber \\
  \Sigma_{i,j} & = t_i + t_j,\nonumber                       \\
  \Delta_{i,j} & = t_i - t_j,\nonumber                       \\
  \Pi_{i,j}    & = t_i t_j.
\end{align}
We follow \citet{baluev_assessing_2008} and neglect aliasing effects.
In this approximation all the terms containing sine or cosine of $\nu \Sigma$ average out.
The terms containing $\sin\nu\Delta$ can also be neglected.
Indeed, in the low frequency limit ($\nu\Delta \ll 1$),
the terms in $\sin\nu\Delta$ vanish,
while in the high frequency limit ($\nu\Delta \gg 1$), the terms in $\sin\nu\Delta$ average out.
We thus obtain
\begin{align}
  \label{eq:QSRnoalias}
  Q & \approx \frac{1}{2}\varmean{\cos\nu\Delta} \id,\nonumber             \\
  S & \approx \frac{1}{4}\varmean{\Sigma\hadprod\cos\nu\Delta} J,\nonumber \\
  R & \approx \frac{1}{2} \varmean{\Pi\hadprod\cos\nu\Delta} \id,
\end{align}
where $J$ is the antisymmetric matrix
\begin{equation}
  J=\left(\begin{array}{cc}
    0  & 1 \\
    -1 & 0
  \end{array}\right).
\end{equation}
As in \citet{baluev_assessing_2008},
we additionally assume that $\psi(\nu)$ is orthogonal to $\psi_\H$ for all $\nu$.
As a consequence, $\tilde{Q}=Q$, $\tilde{S}=S$, $\tilde{R}=R$, and
\begin{equation}
  \label{eq:Mortho}
  M(\nu)  = Q^{-1}\left(R - S\t Q^{-1}S\right).
\end{equation}
Replacing Eq.~(\ref{eq:QSRnoalias}) in Eq.~(\ref{eq:Mortho}), we find
\begin{equation}
  M(\nu) \approx \left(
  \frac{\varmean{\Pi\hadprod\cos\nu\Delta}}{\varmean{\cos\nu\Delta}}
  - \left(\frac{\varmean{\Sigma\hadprod\cos\nu\Delta}}{2\varmean{\cos\nu\Delta}}\right)^2\right)\id.
\end{equation}

The two eigenvalues of $M$ are thus equal in this approximation, and
we can approximate $T_\mathrm{eff}$ with (see Eq.~(\ref{eq:Teff}))
\begin{equation}
  \label{eq:Teff2}
  T_\mathrm{eff} \approx \sqrt{4\pi}\mean{\sqrt{\frac{\varmean{\Pi\hadprod\cos\nu\Delta}}{\varmean{\cos\nu\Delta}}
      - \left(\frac{\varmean{\Sigma\hadprod\cos\nu\Delta}}{2{\varmean{\cos\nu\Delta}}}\right)^2}}.
\end{equation}
In the low frequency limit $\nu_\mathrm{max}\Delta \ll 1$, the cosines can all be replaced by 1,
and we obtain
\begin{align}
  \label{eq:Tefflow}
  T_\mathrm{eff,\,low} & \approx \sqrt{4\pi}\sqrt{\frac{\varmean{\Pi}}{\varmean{1}}
    - \left(\frac{\varmean{\Sigma}}{2\varmean{1}}\right)^2}\nonumber\\
  & \approx \sqrt{4\pi} \sqrt{\frac{t\t C^{-1} t}{u\t C^{-1}u} - \left(\frac{u\t C^{-1} t}{u\t C^{-1}u}\right)^2},
\end{align}
where $u$ is the vector of size $n$ filled with ones.
This expression was proposed by \citet{baluev_planetpack_2013} as a "suggestive generalization"
of the results of \citet{baluev_assessing_2008} to correlated noise.
However, \citet{baluev_planetpack_2013} advocates against its use since it was not proved rigorously.
Moreover, this expression is not valid in the general case but only in the low frequency limit.

In the high frequency limit $\nu_\mathrm{max}\Delta \gg 1$, the cosines average out in Eq.~(\ref{eq:Teff2}),
except on the diagonal (where they are equal to 1), and we find
\begin{equation}
  \label{eq:Teffhigh}
  T_\mathrm{eff,\,high} \approx \sqrt{4\pi}
  \sqrt{\frac{w\t(t\hadprod t)}{w\t u} - \left(\frac{w\t t}{w\t u}\right)^2},
\end{equation}
where $w = \diag(C^{-1})$.

Finally, for any frequency $\nu_\mathrm{max}$, we can approximate (at first order)
the integral of Eq.~(\ref{eq:Teff2})
by replacing each sum $\varmean{X(\nu)}$ by its average $\mean{\varmean{X}}$
over the range $]0,\nu_\mathrm{max}]$.
We have
\begin{align}
  \mean{\varmean{\cos\nu\Delta}}               & = \varmean{\sinc \nu_\mathrm{max}\Delta},\nonumber               \\
  \mean{\varmean{\Sigma\hadprod\cos\nu\Delta}} & = \varmean{\Sigma\hadprod\sinc \nu_\mathrm{max}\Delta},\nonumber \\
  \mean{\varmean{\Pi\hadprod\cos\nu\Delta}}    & = \varmean{\Pi\hadprod\sinc \nu_\mathrm{max}\Delta},
\end{align}
and thus
\begin{equation}
  \label{eq:Teff3}
  T_\mathrm{eff} \approx \sqrt{4\pi}
  \sqrt{\frac{\varmean{\Pi\hadprod\sinc \nu_\mathrm{max}\Delta}}{\varmean{\sinc \nu_\mathrm{max}\Delta}}
    - \left(\frac{\varmean{\Sigma\hadprod\sinc \nu_\mathrm{max}\Delta}}{2{\varmean{\sinc \nu_\mathrm{max}\Delta}}}\right)^2}.
\end{equation}
Equations~(\ref{eq:Tefflow}) and (\ref{eq:Teffhigh}) can also be derived directly from Eq.~(\ref{eq:Teff3})
in the low frequency and high frequency approximations.
Moreover, as explained in Sect.~\ref{sec:fap},
the expression found by \citet{baluev_assessing_2008} in the white noise case
can also be derived from Eq.~\ref{eq:Teff3} by using the fact that the covariance matrix is diagonal.

In practical applications, all estimations of $T_\mathrm{eff}$
(numerical evaluation of Eqs.~(\ref{eq:Teff}) or (\ref{eq:Teff2}), or directly using
Eqs.~(\ref{eq:Tefflow}),~(\ref{eq:Teffhigh}), and (\ref{eq:Teff3})) yield similar results.
Moreover, as noted by \citet{baluev_assessing_2008} in the case of independent Gaussian noise,
$T_\mathrm{eff}$ is often of the same order of magnitude
as the total time span of the time series ($T_\mathrm{span} = \max(t)-\min(t)$).

\section{Periodogram expectation}
\label{sec:wrongnoiseformula}

In this appendix, we show how to obtain an analytical estimate of the expectation of
a periodogram computed with an incorrect noise model.
We assume that the actual covariance matrix $C$ of the noise is not known
and that the $\chi^2$ and periodograms are computed using an incorrect
covariance matrix $V$.
Under the null hypothesis (model $\H$), the time series still follows
Eq.~(\ref{eq:yH}) but the $\chi^2$ becomes
\begin{align}
  \chi^2(\theta_m) & = (y-\varphi_m\theta_m)\t V^{-1} (y-\varphi_m\theta_m)\nonumber      \\
                   & = (\varphi_\H \theta_{\H,0} - \varphi_m\theta_m + \epsilon)\t V^{-1}
  (\varphi_\H \theta_{\H,0} - \varphi_m\theta_m + \epsilon).
\end{align}
The least squares estimate of the parameters becomes
\begin{equation}
  \hat{\theta}_m = \theta_{\H,0} + \left(\varphi_m\t V^{-1} \varphi_m\right)^{-1} \varphi_m\t V^{-1} \epsilon,
\end{equation}
and the minimum $\chi^2$ is thus
\begin{equation}
  \label{eq:chi2eps}
  \chi^2_m = \epsilon\t \left(V^{-1} - V^{-1} \varphi_m \left(\varphi_m\t V^{-1} \varphi_m\right)^{-1} \varphi_m\t V^{-1}\right) \epsilon.
\end{equation}
We introduce a change of coordinates that is slightly different
from Sect.~\ref{sec:changevar} (Eq.~(\ref{eq:changevar})), i.e.,
\begin{align}
  \eta   & = L^{-1} \epsilon,\nonumber\\
  \zeta_m & = M^{-1} \varphi_m,\nonumber\\
  N & = M^{-1} L,
\end{align}
where $C=L L\t$ and $V = M M\t$ are the Cholesky decompositions
of the true and assumed covariance matrices,
and $\eta$ is a vector of independent, centered, and reduced Gaussian random variables.
In these coordinates, the minimum $\chi^2$ (Eq.~(\ref{eq:chi2eps})) is rewritten as
\begin{equation}
  \chi^2_m = \eta\t N\t \left(\id - P_{V,m}\right) N \eta,
\end{equation}
where
\begin{equation}
  P_{V,m} = \zeta_m \left(\zeta_m\t\zeta_m\right)^{-1} \zeta_m\t
\end{equation}
is the projection matrix on the subspace of $\R^n$ defined by the vectors of $\zeta_m$.
The expectation of the minimum $\chi^2$ with the wrong covariance matrix $V$ is thus
\begin{align}
  \label{eq:mum}
  \mu_m &= \E(\chi^2_m) = \tr\left(N\t \left(\id - P_{V,m}\right) N\right)\nonumber\\
  &= \tr\left(\left(\id - P_{V,m}\right) C_V\right),
\end{align}
where $C_V = N N\t = M^{-1} C M\mt$.
In the case $V = C$, we have $C_V = \id$, and we deduce
\begin{align}
  \mu_\H &= n_\H,\nonumber\\
  \mu_\K &= n_\K.
\end{align}

\subsection{First order formula}

At first order, the expectation of the periodogram
can be obtained by replacing $\chi^2_\H$ (respectively, $\chi^2_\K$)
by its expectation $\mu_\H$ (respectively, $\mu_\K$)
in the definition of the periodogram (Eq.~(\ref{eq:defperio})).
We find
\begin{align}
  \label{eq:expectperioO1}
  \E(z_0(\nu)) = \frac{1}{2}\left(\mu_\H-\mu_\K(\nu)\right), \qquad &
  \E(z_1(\nu)) \approx \frac{n_\H}{2}\frac{\mu_\H-\mu_\K(\nu)}{\mu_\H},\nonumber\\
  \E(z_2(\nu)) \approx \frac{n_\K}{2}\frac{\mu_\H-\mu_\K(\nu)}{\mu_\K(\nu)}, \qquad &
  \E(z_3(\nu)) \approx \frac{n_\K}{2}\ln\frac{\mu_\H}{\mu_\K(\nu)}.
\end{align}
In the case $V=C$ (the actual covariance matrix is known), we deduce
\begin{equation}
  \label{eq:expectperiotrue}
  \E(z_i(\nu)) \approx \frac{d}{2},
\end{equation}
for $i =0,\dots,3$, and for all frequencies $\nu$.
However, in the case $V\neq C$ (wrong noise model),
the periodogram expectation can significantly depart from $d/2$
and depends on the frequency $\nu$.

\subsection{Higher order formulas}

Higher order estimates can also be obtained
by developing Eq.~(\ref{eq:defperio}) in power series of $\chi^2_m-\mu_m$
and computing higher order momenta of $\chi^2_\H$, $\chi^2_\K$.
We provide more details in the following,
however, the first order formula already yields very accurate results,
and we thus adopt it in our study.

We introduce the random variables $X_m = \chi^2_m - \mu_m$ ($m=\H,\ \K$),
which we assume to be small with respect to $\mu_m$.
We then develop the periodogram power of Eq.~(\ref{eq:defperio}) in power series of $X_m$.
For instance, at second order we obtain
\begin{align}
  \label{eq:devperio}
  z_0(\nu) &= \frac{1}{2}\left(\mu_\H-\mu_\K(\nu) + X_\H - X_\K(\nu)\right),\nonumber\\
  z_1(\nu) &= \frac{n_\H}{2} \left(1-\frac{\mu_\K(\nu)}{\mu_\H}
    \left(1+\frac{X_\K(\nu)}{\mu_\K(\nu)}
      -\frac{X_\H}{\mu_\H}
      \right.\right.\nonumber\\
      &\hspace{2.7cm}\left.\left.
      -\frac{X_\H X_\K(\nu)}{\mu_\H\mu_\K(\nu)}
      +\left(\frac{X_\H}{\mu_\H}\right)^2
      \right)\right) + \O\left(X^3\right),\nonumber\\
  z_2(\nu) &= \frac{n_\K}{2} \left(-1+\frac{\mu_\H}{\mu_\K(\nu)}
    \left(1+\frac{X_\H}{\mu_\H}
    -\frac{X_\K(\nu)}{\mu_\K(\nu)}
      \right.\right.\nonumber\\
      &\hspace{2.9cm}\left.\left.
      -\frac{X_\H X_\K(\nu)}{\mu_\H\mu_\K(\nu)}
      +\left(\frac{X_\K(\nu)}{\mu_\K(\nu)}\right)^2
      \right)\right) + \O\left(X^3\right),\nonumber\\
  z_3(\nu) &= \frac{n_\K }{2} \left(
    \ln\frac{\mu_\H}{\mu_\K(\nu)}
    + \frac{X_\H}{\mu_\H} - \frac{X_\K(\nu)}{\mu_\K(\nu)}
    \right.\nonumber\\
    &\hspace{1.0cm}\left.
    - \frac{1}{2}\left(\frac{X_\H}{\mu_\H}\right)^2
    + \frac{1}{2}\left(\frac{X_\K(\nu)}{\mu_\K(\nu)}\right)^2
  \right) + \O\left(X^3\right).
\end{align}
The expectation of the periodogram is thus (at second order)
\begin{align}
  \label{eq:expectperioO2}
  \E(z_0(\nu)) &= \frac{1}{2}\left(\mu_\H-\mu_\K(\nu)\right),\nonumber\\
  \E(z_1(\nu)) &\approx \frac{n_\H}{2}\left(1-\frac{\mu_\K(\nu)}{\mu_\H}
    + \frac{\cov(\chi^2_\H,\chi^2_\K(\nu))}{\mu_\H^2} - \frac{\var(\chi^2_\H) \mu_\K(\nu)}{\mu_\H^3} \right),\nonumber\\
  \E(z_2(\nu)) &\approx \frac{n_\K}{2}\left(\frac{\mu_\H}{\mu_\K(\nu)} - 1
    + \frac{\var(\chi^2_\K(\nu)) \mu_\H}{\mu_\K(\nu)^3} - \frac{\cov(\chi^2_\H,\chi^2_\K(\nu))}{\mu_\K(\nu)^2}\right),\nonumber\\
  \E(z_3(\nu)) &\approx \frac{n_\K}{2}\left(\ln\frac{\mu_\H}{\mu_\K(\nu)}
    + \frac{\var(\chi^2_\K(\nu))}{2\mu_\K(\nu)^2} - \frac{\var(\chi^2_\H)}{2\mu_\H^2}\right),
\end{align}
where $\mu_\H$, $\mu_\K(\nu)$ are computed according to Eq.~(\ref{eq:mum}),
and
\begin{align}
  \label{eq:varm}
  \cov(\chi^2_m, \chi^2_{m'})
  & = \tr\left(N\t \left(\id - P_{V,m}\right) N N\t \left(\id - P_{V,m'}\right) N\right)\nonumber\\
  & = \tr\left(\left(\id - P_{V,m}\right) C_V \left(\id - P_{V,m'}\right) C_V\right).
\end{align}

\end{document}